# Optical performance of a new design of trifocal intraocular lens based on the Devil's diffractive lens

Walter D. Furlan[1], Anabel Martínez-Espert[1,*], Diego Montagud-Martínez[1,2], Vicente Ferrando[2], Salvador García-Delpech[3], and Juan A. Monsoriu[2]

*Corresponding author: anabel.martinez@uv.es

[1]Departamento de Óptica y Optometría y Ciencias de la Visión, Universitat de València, Burjassot, Spain.
[2]Centro de Tecnologías Físicas, Universitat Politècnica de València, Valencia, Spain.
[3]Clínica Aiken, Fundación Aiken, Valencia, Spain.

## Abstract

In this work, we propose a new diffractive trifocal intraocular lens design with focus extension, conceived to provide a high visual performance at intermediate distances. This design is based on a fractal structure known as the "Devil's staircase". To assess its optical performance, numerical simulations have been performed with a ray tracing program using the Liou-Brennan model eye under polychromatic illumination. The simulated through the focus visual acuity was the merit function employed to test its pupil-dependence and its behavior against decentering. A qualitative assessment of the multifocal intraocular lens (MIOL) was also performed experimentally with an adaptive optics visual simulator. The experimental results confirm our numerical predictions. We found that our MIOL design has a trifocal profile, which is very robust to decentration and has low degree of pupil dependence. It performs better at intermediate distances than at near distances and, for a pupil diameter of 3 mm, it works like an EDoF lens over almost the entire defocus range.

## Introduction

The increase in life expectancy of the population in recent decades has led to an increase in the number of presbyopic patients with ever more demanding visual requirements [1, 2]. Due to the emergence of new technologies such as tablets and smartphones, and the use of computers in almost all occupational environments [3, 4] these demands are mainly focused on improving intermediate distance vision.

One of the surgical treatments for presbyopia is the replacement of the crystalline lens with a multifocal intraocular lens (MIOL). Although historically, this surgery has been performed after the onset of cataracts, nowadays it is increasingly common to replace the crystalline lens with a MIOL in presbyopic patients, even before it becomes opacified. However, not all MIOLs available on the market are suitable to meet the above-mentioned demands, since bifocal intraocular lenses (IOLs) lack a focus for intermediate vision and trifocal IOLs also have limited performance at these distances [5, 6]. Therefore, extended depth of focus (EDoF) lenses have been proposed more recently to widen the distance focus and to improve visual performance in intermediate vision. However, the major limitation of this type of IOLs is their low performance in near vision [7, 8]. Some other limitations of current MIOLs are the photic phenomena such as halos and glare they generate, and their dependence on pupil dynamics and centration, especially in designs with different power zones or aspheric optics.

In order to create new MIOL designs that overcome the shortcomings of current models, i.e., a MIOL design that in addition to improving intermediate vision performance will not limit near vision, it must be taken into account that, the optical performance may be different under different illumination conditions. Therefore, when designing a MIOL, it is necessary to know what the optical performance is, not only for the MIOL design wavelength, but also for other wavelengths and fundamentally for polychromatic light, since these are the real-life conditions [9, 10].

In this paper we propose a new diffractive trifocal MIOL design with focus extension, conceived to improve visual performance at intermediate distances. This design is based on a fractal structure known as the "Devil's staircase" [11]. Thanks to its unique focusing properties, this type of diffractive lens has been successfully used in different applications such as optical tweezers [12, 13], optical vortices [14] and optical encryption [15]. In addition, in visual optics, our research group has applied this new technology to propose contact lenses for myopia control [16, 17] and a bifocal IOL with focus extension, both in the near and far vision focus [18-20]. Following this line of research, the first results of a trifocal IOL based on the same principle are presented in this work. For its theoretical study, numerical simulations have been performed with a ray tracing program using the Liou-Brennan model eye. To obtain quantitative information on the expected clinical outcomes after IOL implantation, the area under the modulation transfer function (MTFa) was used as merit function, since it has been shown that it is highly correlated with the clinical visual acuity (VA) [21-23]. The optical performance including its pupil-dependence and its behavior against decentering has been assessed with polychromatic light. Finally, to obtain a qualitative assessment of the IOL behavior, it was evaluated experimentally with an adaptive optics visual simulator.

## Methods

### Lens design

The proposed MIOLs was conceived as a hybrid system, in which a refractive IOL of a given base power is combined with a diffractive lens, known as Devil's Lens [11]. The design of a Devil's lens is based on the Cantor function or Devil's staircase function [24], which is defined as

$$F_S(x) = \begin{cases} \dfrac{l}{2^S} & if \quad p_{S,l} \leq x \leq q_{S,l} \\ \dfrac{1}{2^S}\dfrac{x - q_{S,l}}{p_{S,l+1} - q_{S,l}} + \dfrac{l}{2^S} & if \quad q_{S,l} \leq x \leq p_{S,l+1} \end{cases} \quad (1)$$

with $x \in [0,1]$, being $F_S(0) = 0$ and $F_S(1) = 1$. The intervals $[p_{S,l}, q_{S,l}]$, with $l = 1, \ldots, 2^S - 1$, correspond to the gaps limits of the triadic Cantor fractal set of order $S$. Fig. 1 shows the triadic Cantor set developed up to $S = 3$ and the corresponding Cantor function $F_3(x)$. In this example, the triadic fractal Cantor set has $2^S = 8$ segments of length $3^{-S} = 1/27$ and $2^S - 1 = 7$ gaps, where $F_3(x)$ takes the constant values $l/8$, with $l = 1, \ldots, 7$, and increases linearly between these intervals.

As can be seen in Fig. 1 as result of the triadic Cantor set the Devil's staircase function maps the interval [0,1] onto [0,1], and it is constant almost everywhere except for the values of the Cantor set. From the generating function $F_S(x)$ we define the corresponding generalized Devil's lens as a pure phase circularly symmetric diffractive optical element whose transmittance is given by

$$q(\zeta) = exp[-i \; \pi \; 2^{S+1} \, K \, F_S(\zeta)], \quad (2)$$

where $K$ is a positive integer number and $\zeta = (r/b)^2$ is the normalized quadratic radial variable being $b$ is the radius of the lens. Thus, the phase variation along the radial coordinate is quadratic in each zone of the lens. Note that the phase shift at the gap regions is $-l2\pi K$, with $l = 1, \ldots, 2^S - 1$.

It has been demonstrated that under plane wave illumination a Devil's lens produces a focal volume with a characteristic fractal profile around a main focal point [11]. In this case the focal point is located at a distance $f_S = b^2/(2\lambda_0 K3^S)$, where $\lambda_0$ is the design wavelength. Thus, a Devil's lens has four design parameters (namely: $b$, $K$, $S$ and $\lambda_0$) that can be adjusted to obtain a given main focal distance. It is important to note that even with a fixed value of $f_S$, different combinations these parameters allow to obtain very different focusing behaviors. In fact, in previous papers we have proposed the first IOLs with a fractal structure [18-20] in which we used the values $S = 2, K = 3$ and $b = 2.9$ mm. At that time our aim was to obtain bifocal IOLs with an extension of the depth of focus. Actually, as a consequence of the fractality of the lens, with these values of *S* and *K* we obtained two complimentary designs. One was a center-distance EDOF design that provides a clear dominance of the far focus over the near focus [20]; and the other one was a center-near EDoF design that produces just the opposite effect [19].

In this work, we exploited the versatility of the Devil's lenses design to obtain the first trifocal-fractal IOL. The IOL profile was obtained as follows: starting from a spherical refractive monofocal IOL of an arbitrary power, we act on one of the surfaces by adding the diffractive profile $q(\zeta)$ (see Fig. 2).

The new Devil's MIOL has been designed for a wavelength of 550 nm with a refractive index of 1.46. The following values: $S = 3, K = 1$, and $b = 2.9$ mm, were employed in order to obtain the near focus of the MIOLs which is provided by the first diffractive order Devil's lens. Therefore, the IOL nominal addition (*Add*), i.e., the difference between the near and far powers results $Ad = 1/f_S = 2\lambda_0 K3^S/b^2 = 3.5$ D. With these values a lower addition power of amount 1.75 D is also obtained, which is created by the cluster of secondary foci generated by the fractal structure [11]. In this way, we leave the zeroth order of diffraction for the far focus which, consequently, is purely refractive.

## *Numerical evaluation*

The optical properties of the Devil's MIOL design were initially performed numerically with Zemax OpticStudio software (v. 18.7, LLC, Kirkland, WA, USA). The Liou–Brennan model eye [25] was employed, but the crystalline lens of the model was replaced by the MIOL design. The study was conducted for 3.0 mm and 4.5 mm pupil diameters (emulating photopic and mesopic conditions). The diffractive profile of the Devil's MIOL surface (see Fig. 1) was inserted as a Grid Sag Surface in the first surface of a refractive lens of 20 D. An aspheric surface was chosen as the back surface to neutralize the total spherical aberration of the model eye. Table 1 shows the parameters of the model eye and the MIOL.

| Surface | Radius of curvature (mm) | Asphericity | Thickness (mm) | Refractive index | Abbe number |
|---|---|---|---|---|---|
| Anterior Cornea | 7.77 | -0.18 | 0.50 | 1.376 | 50.23 |
| Posterior Cornea | 6.40 | -0.60 | 3.16 | 1.336 | 50.23 |
| Pupil | - | - | 0.00 | 1.336 | 50.23 |
| Anterior IOL Diffractive profile | 21.40 | 0.00 | 0.70 | 1.46 | 58.00 |
| Posterior IOL aspheric | -17.75 | -32 | 18.65 | 1.336 | 50.23 |

Table 1. Parameters of the model eye with the MIOL.

First, the modulation transfer functions (MTFs) were obtained for vergences between -0.50 D and +3.50 D in steps of 0.10 D. Then, the MTFa were calculated for frequencies between 0 lines/mm and 50 lines/mm. These functions were calculated for 450 nm, 550 nm and 650 nm wavelengths, and also with polychromatic light using the V(λ) function given by the software. Using these polychromatic MTFa, we have computed the VA defocus curves that could be obtained with the implanted MIOL. To this end, we have employed the following semi-empirical equation [21]

$$VA = 1.828 * e^{-0.230*MTF_a} + 0.014 \qquad (3)$$

because, using this expression, Armengol et al. [21], found a high correlation ($R^2 = 0.94$) between VA measured in patients implanted with commercial MIOLs versus the polychromatic MTFa obtained on an optical bench for the same lenses.

In addition, taking into account that, depending on the MIOL design, a decentration may affect its optical performance to a greater or lesser extent [26], we also calculated the VA defocus curves to test the consequences of a lens decentration. Finally, following the target of predicting clinical results with this new design, simulations of the images that would be formed in the retina were obtained from the convolution of an optotype with the numerical point spread functions (PSFs), calculated at the main foci with polychromatic light.

### *Experimental evaluation*

To obtain qualitative results of the visual performance of the new MIOL design, experimental measurements were performed using an adaptive optics visual simulator (VAO, Voptica SL, Murcia, Spain) [27]. This instrument allows simulating the phase profile of an IOL by projecting it onto the pupil plane of the eye under test. It can also simulate different amounts of defocus with which a defocus curve can be obtained with the patient's VA. In our case, in order to obtain more objective experimental measurements, a camera was attached to the commercial device, which acted as an artificial presbyopic eye. Images of a letter optotype at different vergences were obtained through the phase profile of the MIOL. The camera has a CMOS sensor (EO-10012C LE, 8-bit, 3840 x 2747 pixels, 6.41 mm x 4.59 mm) and an achromatic doublet lens (AC254-030-A-ML, Thorlabs Inc., Newton, NJ, USA). In addition, since the original commercial device has a single 4.5 mm pupil by default, an additional pupil diameter of 3.0 mm has been incorporated. In this way it was possible to make a rapid objective assessment of the MIOL visual performance measurement without having to involve real patients. Measurements were carried out with polychromatic light. In the experiment a high photopic luminance (120 cd/m$^2$) and medium photopic luminance (80 cd/m$^2$) were used for the 3.0 mm pupil and for the 4.5 mm pupil, respectively. An optotype with Snellen Es of sizes corresponding to visual acuities of 0.4 logMAR, 0.2 logMAR and 0.0 logMAR was used as object.

## Results

### *Numerical evaluation*

Figure 3 shows the monochromatic MTFa curves obtained with three wavelengths (450 nm, 550 nm and 650 nm) and with polychromatic light. It is evident that for both pupils in the model eye, the new design produces a trifocal response for the three wavelengths and that the locations of the maximum values for the design wavelength and polychromatic light are coincident and located at 0.0 D (distance), +1.4 D (intermediate) and +2.8 D (near). For both pupils, at the far focus (0.0 D defocus), there is a longitudinal chromatic aberration (LCA) of 1.1 D and the order in which the maxima appear for the three wavelengths confirms the refractive origin of this focus. On the other hand, as expected, it is clearly seen how in the intermediate focus and especially in the near focus the LCA values are partially compensated by the diffractive effect of the IOL.

As a consequence of the LCA, the maximum MTFa value (0.0 D defocus) for polychromatic light is reduced with respect to the MTFa obtained for the design wavelength. This reduction is approximately of 15% and 26%, for 3.0 mm pupil and for 4.5 mm pupil respectively. On the other hand, if we compare the heights of these curves (black and green) at the intermediate and near foci, they are practically the same for both pupils.

The differences between the MTFas curves obtained for the 3.0 mm and 4.5 mm pupils indicate that the MIOL shows some pupil-dependence. This aspect can be further analyzed in Figure 4 in which the polychromatic MTFa values for the three main foci are plotted as a function of pupil diameter in steps of 0.25 mm. It can be seen that the MTFa is nearly constant for three foci of the MIOL up to a pupillary diameter of 3.5 mm. From 3.5 to 4.5 mm, as the diameter increases the MTFa decreases slightly for the intermediate and near foci and increases for the distance focus. Note, that for the whole range of pupillary diameters in Fig. 4 the intermediate vision focus is always better than the near vision focus.

Figure 5 shows the VA defocus curves, obtained from the polychromatic MTFa with Eq. (3), and the images at the main foci simulated in both cases with the MIOL centered on the visual axis (solid black line) and decentered 0.25 mm, a typical value [28], toward the temporal side (dashed magenta line).

As can be seen in Fig. 5 a), the VA defocus curves obtained for both pupils with the centered and decentered MIOL have no significant differences. On the other hand, as with the MTFa curves, the VA variations throughout the defocus range are smaller for the 3.0 mm pupil than for the 4.5 mm pupil. Moreover, for the small pupil these values always remain above 0.1 logMAR between the near and intermediate foci and do not drop below 0.2 logMAR over the entire defocus range between infinity and 33 cm. On the other hand, for the 4.5 mm pupil the far focus achieves a somewhat better VA than for the 3.0 mm pupil but is worse for the intermediate and near foci. However, even for large pupil, the VAs are better than 0.4 logMAR at the 3 main foci.

Figure 5 b) shows simulations of images of a letter optotype obtained at the 3 main foci of the MIOL. The lines framing each image correspond to those in Fig. 5 a). As can be seen in these images there is a clear correspondence with the VA predictions obtained with Eq. (3); i.e., for both pupils, the differences between the centered and decentered images are not appreciable. Moreover, in this case, for the 3.0 mm pupil it is possible to resolve the letter E corresponding to VA 0.0 logMAR in the images corresponding to three main foci. For the 4.5 mm pupil, a decrease in contrast can be seen in the intermediate and near images.

*Experimental results*

Figure 6 shows the images provided by the new trifocal MIOL design obtained with the artificial eye through the VAO visual simulator at the main foci of the lens. For both pupils at the far, intermediate and near foci, it is possible to resolve the line of VA corresponding to 0.0 logMAR. In the images it is also observed that the contrast of the letters decreases as the test approaches to the eye. Although these images are qualitative experimental measurements, they agree well with the quantitative numerical results shown in Fig. 5 b).

## Discussion

In this work we have presented a new trifocal MIOL design based on the diffractive Devil's lens. We found that the Devil's MIOL design with a nominal *Add* value of +3.5 D has a trifocal profile, with the main foci at 0.0, +1.4, and +2.8 D (Fig. 3 and 5). The effective addition power was computed as the dioptric difference between the distance and near inflection points of the defocus curve. The difference between the nominal and effective addition values are in accordance with clinical results recently reported [29]. In fact, Law et al. found that for a MIOL with nominal *Add* of 3.5 D, the mean effective addition predicted with six different biometry formulas was $2.60 \pm 0.29$ D.

From our polychromatic study (Fig. 3) it is observed that regardless of pupillary size, in the far vision the LCA values are approximately 1.1 D, which are within the values measured in healthy phakic eyes [30]. That is, the proposed model would not induce abnormal LCA at the far focus. Moreover, the LCA at the intermediate and near foci would be reduced due to the chromatic aberration of the diffractive part of the MIOL.

The results in Fig. 4 show a low degree of pupil dependence, with a predominance of distance vision over intermediate vision and of intermediate vision over near. Note, that in our design a large pupil favors even more distance vision, which would be a benefit in mesopic vision, especially for night drivers.

From the defocus curves of VA (Fig. 5) it can be seen that, for small pupil, the proposed lens has the potential to provide VA values equal to or better than 0.2 logMAR, over the whole range of vision, where the intermediate focus always has a higher optical performance than the near focus (Fig. 5). In this sense, if we consider the definition of EDoF lens, from The American Academy of Ophthalmology working group "An EDoF is a lens that provides an extension of focus that is at least 0.50 D wider than for a monofocal IOL at a visual acuity of 0.2 logMAR" [31]. In photopic vision, the proposed MIOL works as an EDoF lens in almost all the defocus range that would meet the needs of patients, without interfering with its best performance for distance in mesopic vision. On the other hand, compared to other commercial trifocal designs, this new design has a reduced number of diffractive rings. This reduction in the number of rings could be related to a potential lower incidence of dysphotopsias [32].

On the other hand, it is known that decentration of an IOL can have negative consequences on the visual performance of the implanted eye. These decentrations have been associated with numerous factors, including the IOL design itself [28, 33]. Therefore, in this study, the performance of the decentration lens has been evaluated. The results (Fig. 5) show that our design is very robust to decentration since there is almost no change between the curves obtained with the centered and decentered lens.

The experimental results obtained with the adaptive optics visual simulator (Fig. 6) are in good agreement with the numerical results obtained with the ray tracing program (Fig. 5). A pre-clinical evaluation of proposed trifocal MIOL with real patients using the VAO system is now in progress.

Finally, it is important to note that the diffractive design could itself constitute a supplementary IOL designed specifically for implantation in the ciliary sulcus. Thus, it could be used to create multifocality in already pseudophakic eyes and even correct postoperative refractive defects without the need to change the monofocal IOL previously implanted in the capsular bag.

Based on the above, we can conclude that our trifocal lens design, created from the Devil's lens structure, is a design that has the potential to overcome the shortcomings of some commercial lenses, especially in the intermediate focus of vision.

Some limitations of this study will be addressed in the future: Some prototypes of the new MIOL will be constructed and further studies should involve in vitro optical quality measurements of the MIOL, in particular the effect on the merit functions (PSF, MTF, and TF-MTF) of the MIOL, Photic phenomena, decentration and tilt.

## Acknowledgments

This work was supported by Ministerio de Ciencia e Innovación, Spain (Grant PID2019-107391RB-I00) and by Generalitat Valenciana, Spain (Grant CIPROM/2022/30). D. M.-M. also acknowledges the Margarita Salas grant from the Ministerio de Universidades, Spain, funded by the European Union-Next Generation EU. A. M.-E. acknowledges financial support from Universitat de València (programa Atracció de Talent 2021).

**Figures**

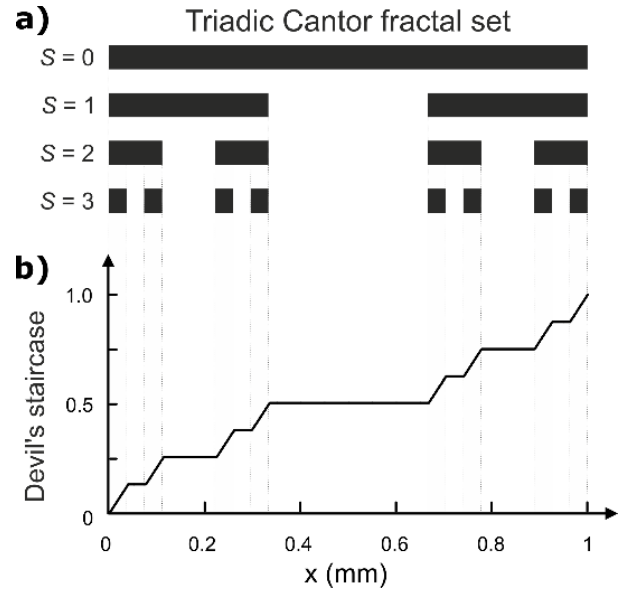

**Figure 1.** a) Triadic Cantor fractal set developed up to $S=3$. The structure for $S=0$ is the initiator and the one corresponding to $S=1$ is the generator constructed by dividing the segment into equal parts of length 1/3 and removing the central one. This procedure is continued at the subsequent stages $S=2$ and 3. b) Cantor function or Devil's staircase, $F_3(x)$.

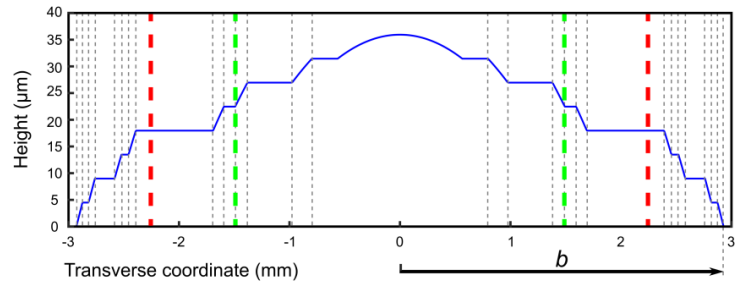

**Figure 2.** Diffractive profile of a Devil's MIOL ($S=3$ and $K=1$) with an addition of 3.5 D. The green and red lines represent pupil radius 3 mm and 4.5 mm, respectively.

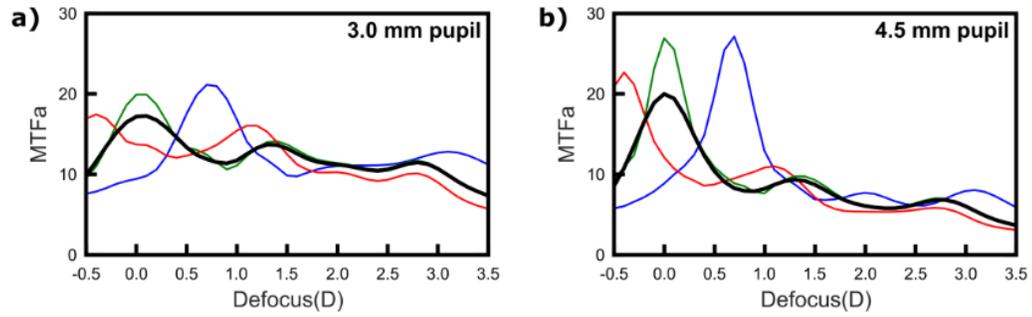

**Figure 3.** MTFa curves for 3.0 mm and 4.5 mm pupils obtained for wavelengths 450 nm (blue line), 550 nm (green line), 650 nm (red line), and polychromatic light (black line).

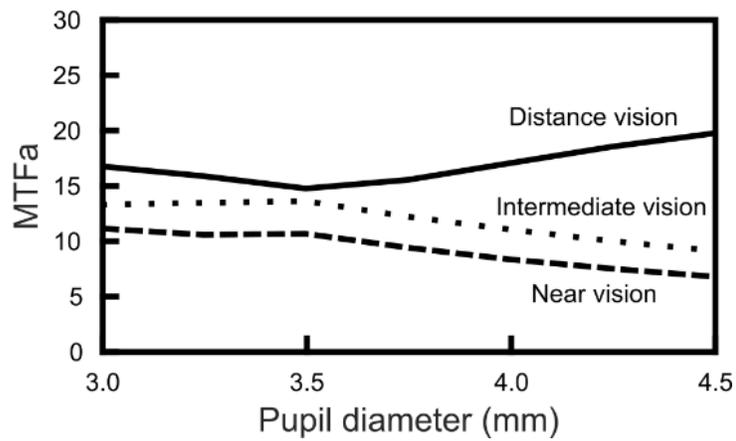

**Figure 4.** MTFa values for the distance, intermediate and near foci obtained for pupils varying from 3.0 mm to 4.5 mm.

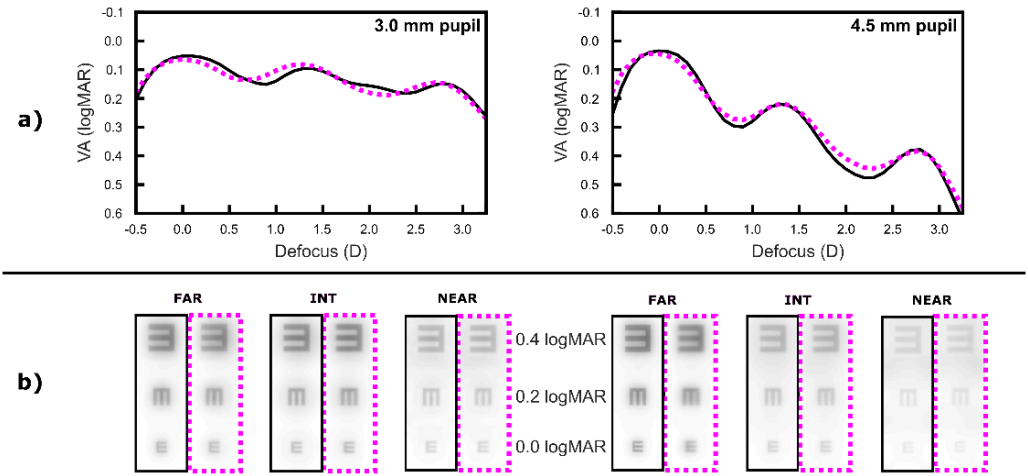

**Figure 5.** a) VA defocus curves with the Devil's MIOL, obtained from polychromatic MTFa values using Eq. (3). The solid black line shows the VA with the lens centered. The dashed magenta line was obtained with the MIOL temporally off-centered 0.25 mm, b) Simulated images obtained (from the PSFs) of an optotype with letters whose sizes correspond to visual acuities of 0.0 logMAR, 0.2 logMAR, and 0.4 logMAR. These images were calculated at the main foci of the MIOL for the centered (black box) and decentered (magenta box) lens.

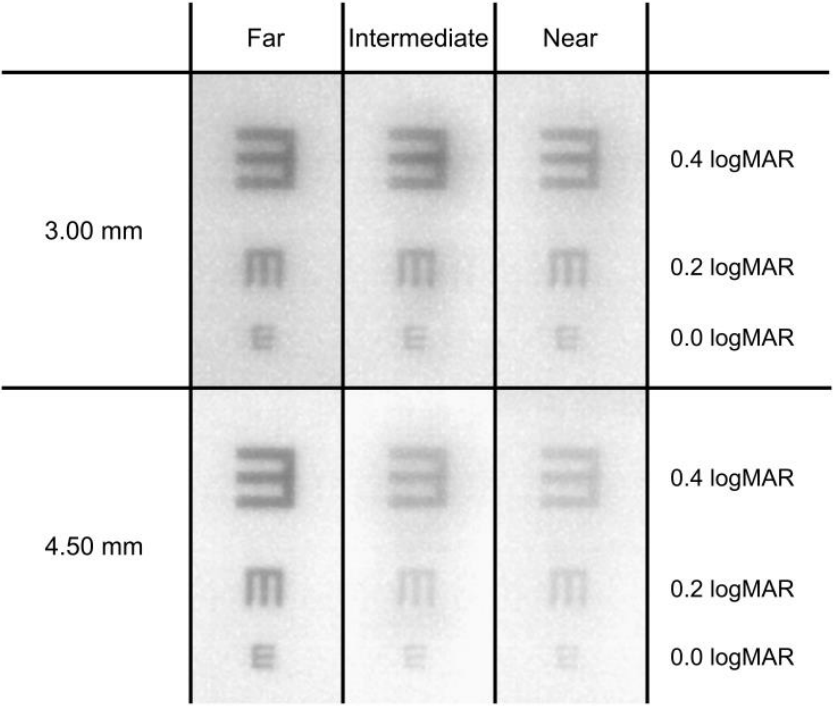

**Figure 6.** Images of a tumbling E optotype corresponding to 0.4, 0.2, and 0.0 logMAR VA obtained with the VAO system simulating the Devil's MIOL.